\documentclass[prb,amsmath,amsymb,floatfix,notitlepage,superscriptaddress,eqsecnum]{revtex4-1}

\usepackage{graphicx}

\usepackage{epstopdf}
\usepackage{amsfonts}
\usepackage{epsfig}

\def\bea{\begin{eqnarray}}
\def\eea{\end{eqnarray}}
\def\be{\begin{equation}}
\def\ee{\end{equation}}

\begin{document}
\title{Screening Clouds and Majorana Fermions\footnote{for  a special issue of J. Stat. Phys. in memory of 
Kenneth G. Wilson}}
\author{Ian Affleck}
 \affiliation{
 Department of Physics and Astronomy, University of British 
Columbia, Vancouver, B.C., Canada, V6T 1Z1}
\author{Domenico Giuliano}
\affiliation{
Dipartimento di Fisica, Universit\`a della Calabria Arcavacata di Rende I-87036, Cosenza, Italy
and
I.N.F.N., Gruppo collegato di Cosenza, Arcavacata di Rende I-87036, Cosenza, Italy}
\affiliation{
CNR – SPIN, Monte S. Angelo, via Cinthia, I-80126, Napoli – Italy}
\date{\today}
\begin{abstract}
Ken Wilson developed the Numerical Renormalization Group technique
which greatly enhanced our understanding of the Kondo effect and other quantum impurity problems. 
Wilson's NRG also inspired Philippe Nozi\`eres to propose the idea of a large ``Kondo screening cloud''.  
While much theoretical evidence has accumulated for this idea it has remained somewhat controversial 
and has not yet been confirmed experimentally. Recently a new possibility  
for observing an analogous crossover  length scale has emerged, involving a Majorana fermion localized at the interface 
between a topological superconductor quantum wire and a normal wire. We give an overview of 
this topic both with and without interactions included in the normal wire.
\end{abstract}
\maketitle
\section{introduction}
A rather unusual feature of  Ken Wilson's contributions to theoretical physics is that they include 
both fundamental conceptual ideas and novel numerical techniques.  The most celebrated 
numerical technique which he helped to develop is lattice gauge theory, now leading to 
a quantitative understanding of the strong interactions. But Wilson also played a huge role 
in condensed matter theory by developing the Numerical Renormalization Group (NRG) technique\cite{Wilson} 
to study the Kondo problem. This technique has since been applied to a host of other quantum 
impurity models and is also used in the Dynamical Mean Field Theory\cite{DMFT} approach to strongly 
correlated systems where translationally invariant models are reduced to impurity models 
by taking the limit of large spatial dimension. 

Typical  quantum impurity models which are studied using the NRG involve localized quantum 
mechanical degrees of freedom, such as a spin-1/2 operator, interacting with a spatially extended 
gapless system such as a Fermi gas of free electrons.  These models can generically be reduced 
to one-dimensional ones, for example by using s-wave projection in cases where the 
impurity interactions are short ranged. Wilson's NRG eventually maps the system 
to a one-dimensional lattice model  with the quantum impurity at one 
end of a chain and the hopping terms between neighbouring sites dropping off exponentially 
with distance from the impurity. These quantum impurity models usually exhibit 
a renormalization group cross-over between ultraviolet and infrared fixed points.  In 
the case of the spin-1/2 Kondo model the ultraviolet fixed point, which describes 
the physics at sufficiently high energies when the bare anti-ferromagnetic Heisenberg exchange coupling 
of the electrons to the spin is very weak, corresponds to a free spin.  The 
infrared fixed point, describing the physics when the renormalized exchange coupling 
is large,  corresponds to a single electron emerging from the Fermi sea 
to form a spin singlet with the impurity. The remaining electrons in the Fermi sea 
must adjust to the presence of this spin singlet, leading to universal low energy behaviour. 
Wilson's NRG approach studies this crossover by focussing on the dependence 
on distance away from the impurity, in particular the crossover as more sites are added to 
the ``Wilson chain''. These ideas led Philippe Nozi\`eres to propose the 
existence of a large ``Kondo screening cloud''.\cite{Nozieres} A cartoon picture is that the 
electron which forms a singlet with the impurity has an extended wave-function of radius 
$\xi_K\approx v_F/T_K$.  Here $T_K$ is the Kondo temperature which is the RG crossover energy scale and 
$v_F$ is the Fermi velocity. We work in units where $k_B=\hbar =1$.  The low 
energy degrees of freedom of the free electrons correspond to relativistic Dirac fermions 
with speed of light replaced by the Fermi velocity, which is thus the natural parameter 
to convert an energy scale, $T_K$ into a length scale, $\xi_K$.

More precisely, one might expect that physical quantities depending on distance 
from the impurity, $r$, should exhibit 
a universal crossover at the length scale $\xi_K$,  being described by the 
ultraviolet fixed point at distances $r\ll \xi_K$ and by the infrared fixed point 
at distances $r\gg \xi_K$.  This naive expectation has been largely confirmed 
by a number of theoretical calculations where RG predictions were compared to 
	NRG and other numerical results.\cite{Kond_rev}   This screening cloud picture is actually 
quite disturbing if one notes that typical values of $\xi_K$ in experiments may correspond 
to hundreds or thousands of lattice spacings. No experiments have yet detected this 
large screening cloud.  A partial explanation of this is that the cloud is 
so big that it is difficult to see. Typical correlation functions involve a product of a 
power law decaying factor, an oscillating factor and a  slowly varying envelope function, 
crossing over at $\xi_K$.  Thus the correlation functions have become very small 
at the distance $\xi_K$ where the crossover occurs. One must also 
consider the fact that in most experiments there is a finite density of quantum impurities 
and their average separations are $\ll \xi_K$. Furthermore, one should consider 
that the underlying models are   indeed just models.  They leave out, for example, 
electron-electron interactions away from the impurity. While this may be 
justifiable based on Fermi liquid theory renormalization group ideas, at 
any finite temperature there is a finite inelastic scattering length beyond 
which ignoring these interactions is not justified. Thus experimental 
confirmation of the Kondo screening cloud  remains elusive.\cite{Kond_rev} 

The existence of a ``screening cloud size'' $\xi_K$, i.e. a characteristic 
length scale for crossover between ultraviolet and infrared fixed points 
in quantum impurity models, is not restricted to the Kondo problem; 
 it is expected to be quite generic.  
  In this article, we will focus on 
 a new type of quantum impurity model where this issue arises 
 and new opportunities present themselves for experimental confirmation. 
 It was pointed out by Kitaev\cite{Kitaev} that a simple 1D p-wave superconductor
defined on a finite line interval has a ``Majorana mode'' localized 
at each end of the system. This is most easily seen for the 
tight-binding model of spinless electrons with equal hopping and pairing amplitudes and zero chemical potential:
\be H=w\sum_{j=1}^{M-1}[-a^\dagger_ja_{j+1}+a_ja_{j+1}+h.c.]\label{Kit}\ee
(Here $h.c.$ stand for Hermitean conjugate and $a_j$ annihilates an electron on lattice site $j$.)
We can always rewrite the $a_j$ operators in terms of their Hermitean and anti-Hermitean parts, so-called Majorana operators:
\be a_j=(\gamma_{2j-1}+i\gamma_{2j})/2,\ \  (j=1,2,3,\ldots M) \ee
where
\be \gamma_j^\dagger =\gamma_j,\ \  \{\gamma_j,\gamma_k\}=2\delta_{j,k}.\ee
$H$ becomes
\be H={iw\over 2}\sum_{j=1}^{M-1}\gamma_{2j}\gamma_{2j+1}.\ee
This model is readily diagonalized by reassembling the Majoranas into ``Dirac'' 
operators in a different way:
\be b_j\equiv (\gamma_{2j}+i\gamma_{2j+1})/2,\ \  (j=1,2,3, \ldots M-1),\ee
yielding:
\be H=w\sum_{j=1}^{M-1}b_j^\dagger b_j.\ee
We obtain $M-1$ single particle levels with energy $w$, which are localized on the links of the lattice. 
However, notice that $\gamma_1$ and $\gamma_{2M}$ don't appear in the Hamiltonian at all.  They 
are a pair of Majorana modes, one at each end of the chain, which will be the focus of our discussion. 
They can, of course, be combined into a Dirac operator giving a zero energy state of the 
chain which has equal amplitude at both ends. 

While the above simple discussion was for a very special model, this phenomenon is actually quite robust, 
persisting for a range of unequal hopping and pairing amplitudes and for a range of chemical potential. 
For more general parameters the Majorana modes (MM's) are localized near the two ends of the chain, 
decaying exponentially towards the centre of the chain with a finite decay length. Thus 
for a long enough chain they are nearly decoupled, leading to an exponentially low energy 
level with equal amplitudes at both ends of the chain. Such a system, containing a MM at each end, is 
known as a topological superconductor. Even allowing general values for the parameters, 
this still might seem like a highly unrealistic model. However, it was shown 
that such a topological phase occurs in a realistic model of a quantum wire with 
spin-orbit interactions, adjacent to a bulk s-wave superconductor, in an applied magnetic field.\cite{Lutchyn,Oreg}
The pairing term then arises from the proximity effect. Signs of these MM's were apparently 
seen in experiments on indium antimonide quantum wires proximity coupled to a 
niobium titanium  nitride superconductor.\cite{Mourik}  In the experiments, one end of the quantum wire extended 
past the edge of the superconductor, over an insulator and then eventually contacted a normal electrode. 
Indications of  a MM localized in the quantum wire near the edge of the superconductor were 
obtained from current measurements. Other experiments were performed in which 
the two ends of the normal wire were in contact with two different superconducting electrodes 
with the central region of the wire resting on an insulator. Curved quantum wires with 
both ends in contact with the same superconducting electrode and a magnetic flux 
applied inside the loop produced by the wire and the superconductor have also been considered 
theoretically. 

\begin{figure}
\includegraphics*[width=0.5\linewidth]{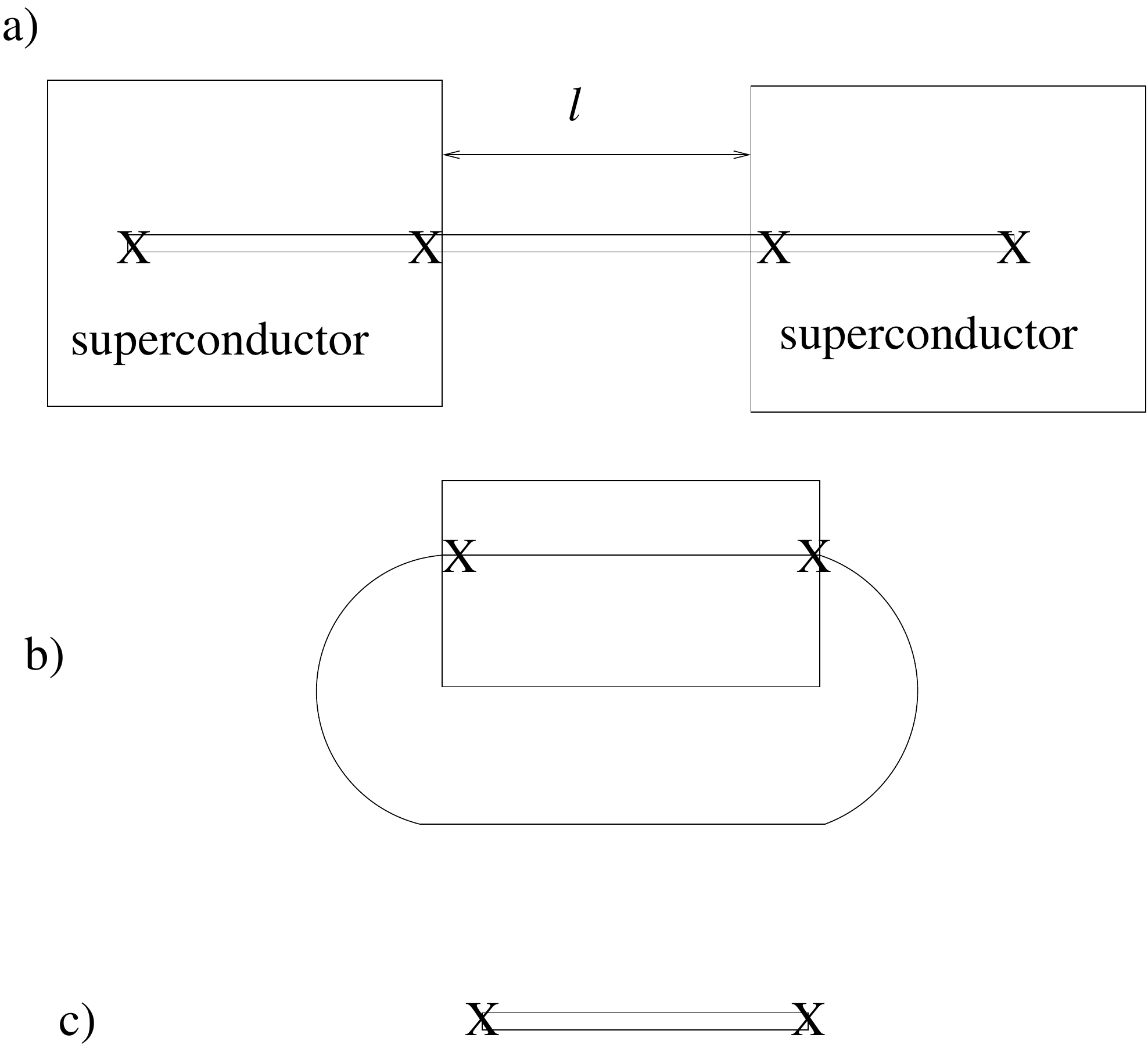}
\caption{a) A sketch of the device considered. The portions of the quantum wire on top of  
the conventional bulk superconductors become topological superconductors, with 
different phases $\chi_L$ and $\chi_R$ of the superconducting order parameter. 
Majorana modes, denoted by black circles, exist at both ends of each topological superconductor. \\
b) A closed curved quantum wire, part of which is on top of a superconductor. Now 
the two Majorana modes of the topological superconductor couple to each end of the normal portion of the wire.\\ 
c) The effective low energy system. Only the normal wire and the 2 Majorana modes 
at the SN junctions are retained.}
 \label{fig:setup}
\end{figure}

We discuss here the low energy behaviour of a long normal wire of length $\ell$ with a MM at each end,\cite{MMs}
corresponding to a Superconductor-Normal-Superconductor (SNS) junction where 
each superconductor is topological.\cite{finite} See Fig.(\ref{fig:setup}).  In order 
to conveniently study the effects of electron-electron interactions inside the 
normal wire and to obtain universal results, we assume that the temperature and the 
finite size gap of the normal wire, ($\propto v_F/\ell$)
are both small compared to the induced superconducting gap in the topological superconductor, $\Delta$.
[$\Delta$ is given by the parameter $w$ in the simple model of Eq. (\ref{Kit}). Note that the 
superconductor is actually gapless, due to the MM. $\Delta$ represents the gap for bulk excitations. 
The MM exists as a single zero energy state inside this gap.] We can then 
safely integrate out the gapped modes in the superconductors, keeping only 
the MM's and the excitations of the normal wire in our low energy effective Hamiltonian, $H_{eff}$.\cite{ACZ,Fidkowski,Affleck}
This leaves a type of 
quantum impurity model, similar to the Kondo model, in which the quantum impurities are 
the MM's and the delocalized gapless excitations are those of the normal wire.
This long junction limit corresponds to $\ell \gg \xi_0$ where $\xi_0\propto v_F/\Delta$ is the 
superconducting coherence length in the topological superconductor. Our assumption $v_F/\ell \ll \Delta$ implies that
there are many Andreev bound states in the normal region. 

 We first 
consider each SN junction separately, appropriate for infinite $\ell$. We show that 
an analogue of Kondo screening occurs.  At low energies,  the MM combines with another 
Majorana degree of freedom within the normal wire to form a Dirac mode localized near the 
SN interface. There is a finite energy cost to depopulate this localized level, 
analogous to the Kondo temperature.  We may think of this localized Dirac mode as being 
extended over some finite distance into the normal wire, $\xi_M$, the analogue of the Kondo 
screening length, $\xi_K$. This state with the MM ``screened'' corresponds to a 
low energy strong coupling fixed point of the model. On the other hand, if the MM is weakly coupled 
to the normal wire, there is an unstable high energy fixed point in which 
the MM is decoupled. 
 We then consider the dc Josephson current through the finite length SNS junction. This 
provides an excellent method for measuring the screening length, $\xi_M$ since the 
current behaves very differently in the two regimes $\xi_M \ll \ell$ and $\xi_M\gg \ell$. Essentially 
the wire length $\ell$ acts as an infrared cut-off on the renormalization of the 
coupling between the MM and the wire. There are strong similarities between this system  
and the 2 impurity Kondo model where a competition occurs between screening 
of both impurities by the electron bath and singlet formation between the 
two impurities due to the Ruderman-Kittel-Kasuya-Yosida (RKKY) interaction.

Note that the MM introduces another characteristic length scale, $\xi_M$, in addition to $\xi_0$ 
the coherence length in the superconductor. For weak tunnelling across the SN junction, 
$\xi_M\gg \xi_0$, the limit we consider. This additional characteristic length scale is a 
special feature of topological superconductor SN junctions, which doesn't exist 
for an ordinary SN junction.  
 While crossover of the Josephson current 
as a function of $\ell /\xi_0$ in both ordinary\cite{Svidzinsky} and topological\cite{Beenakker,Zhang} SNS 
junctions has been frequently analysed,
 the crossover at much longer length 
scales in topological SNS junctions, as a function of $\ell /\xi_M$, is quite novel. This very different 
length dependence in the topological case is discussed further in  Sec. IIA.

There is an important simplification of the MM impurity model compared to the Kondo model. 
The Kondo model exhibits very non-trivial many body effects despite the fact that 
the electron bath is usually treated as non-interacting. These effects are due to the spin degree of freedom of the
impurity and the associated spin-flip scattering. On the other hand, the 
relevant coupling between the MM and the normal wire is a simple tunnelling term 
so that, if we ignore interactions in the normal wire, the Hamiltonian remains harmonic. 
Thus the case of a non-interacting normal wire provides a relatively simple example 
where the screening cloud physics can already be seen. This is a bit reminiscent of the 
non-interacting resonant level model,\cite{Ghosh} which has some relation to the Kondo model although 
in that case the spin degree of freedom of the impurity is dropped. Thus 
we begin by discussing the non-interacting SNS junction.  Simple analytic formulas 
can be obtained for the Josephson current in the two limits $\xi_M \ll \ell$ and $\xi_M\gg \ell$. 
We supplement these with numerical results
which demonstrate the scaling behaviour and the cross-over between weak coupling 
and strong coupling fixed points. We then turn to the case of an interacting normal wire, 
again obtaining analytic expressions for the current at weak and strong coupling fixed points. 
Numerical simulation of the interacting case is more challenging, requiring 
for example use of the Density Matrix Renormalization Group method, and we don't attempt it here.

\section{Non-interacting Model}
We begin by writing down a simple non-interacting tight-binding model. The continuum 
model which captures the universal low energy physics can be derived from this tight-binding 
model which we also find useful for numerical simulations. We decompose the 
Hamiltonian into a ``bulk'' term, $H_0$ describing the normal wire and a boundary 
term, $H_b$ describing the coupling to the two MM's. 
\bea H&=&H_0+H_b \nonumber \\
 H_0&=&- J \sum_{ j = 1}^{N- 2 } \{ c_j^\dagger c_{j+1} + c_{j+1}^\dagger c_j \}
- \mu \sum_{j  = 1}^{N-1} c_j^\dagger c_j\label{H0}\\
H_b&=&- i t_L \gamma_L \{ c_1 e^{ i \frac{\chi_L}{2}} + c_1^\dagger e^{ - i \frac{\chi_L}{2}} \} 
+ t_R \gamma_R  \{ c_{N- 1} e^{ - i \frac{\chi_R}{2}} -  c_{N- 1}^\dagger e^{  i \frac{\chi_R}{2}} \}.\label{Hni}
\eea
Here $\gamma_L$ and $\gamma_R$ are the two MM's manufactured by the two topological superconductors 
at the left and right hand side of the normal wire. $\chi_L$   and $\chi_R$ are the 
phases of the superconducting order parameter in the left and right superconductor. The 
Josephson current is a function of the phase difference $\chi \equiv \chi_L-\chi_R$.  For convenience, 
we henceforth set $\chi_L=\chi /2$ and $\chi_R=-\chi /2$. 
The dc Josephson current is determined by the derivative of the equilibrium free energy, $F$,  
with respect to the phase difference:
\be I[\chi ]=2e{\partial F\over \partial \chi}.\label{Idef}\ee
At $T=0$, which we focus on here, $F$  becomes the ground state energy.

We now turn to a continuum field theory description.  Note that this seems highly 
appropriate since our impurity model is only valid at energy scales below the induced 
superconducting gap of the topological superconductors, a scale which is expected 
to be much less than bandwidth, $J$ of the normal wire.
We also assume 
the tunnelling amplitudes to the MM, $t_{L/R}\ll J$ and that the length of the normal region, $\ell$, is 
large compared to microscopic scales like the lattice constant. Then it is expected that 
the $\chi$ dependence of the ground state energy, and hence the Josephson current, 
depends only on universal low energy information.\cite{Giuliano}
This field theory model 
simplifies calculations in this section and is crucial for including interactions in 
the next section.  

Keeping only a narrow band of wavevectors around the Fermi points, $\pm k_F$, 
we write:
\be c_j\approx \left[ e^{ik_Faj}\psi_+(aj)+e^{-ik_Faj}\psi_-(aj)\right]\sqrt{a}\ee
where $+/-$ label right and left movers respectively, $a$ is the lattice constant and we define $k_F$ by:
 \be \mu = -2J\cos k_Fa.\ee
 The 
fields $\psi_{\pm}$ vary slowly on the lattice scale. Linearizing the dispersion relation 
near the Fermi energy, the low energy bulk Hamiltonian becomes:
\be H_0\approx iv_F\int_0^\ell dx \left[-\psi_+^\dagger {d\over dx}\psi_+ +\psi_-^\dagger {d\over dx}\psi_-\right]
\label{Hnic}\ee
where $v_F=2J\sin k_Fa$.
This Hamiltonian must be supplemented by boundary conditions to be well-defined.  For the lattice 
Hamiltonian of Eq. (\ref{H0}), with free ends, the boundary conditions correspond to requiring the 
operators to vanish on the ``phantom sites'', $j=0$ and $j=N$, implying
	\be 0= \psi_+(0)+\psi_-(0)=e^{ik_F\ell}\psi_+(\ell )+e^{-ik_F\ell}\psi_-(\ell )\label{bc}\ee
where
\be \ell \equiv Na.\ee
It is convenient to make an ``unfolding'' transformation, taking advantage of the boundary condition at $x=0$ to define:
\be \psi_-(-x)\equiv -\psi_+(x),\ \  (0<x<\ell ).\ee
Then we can write the Hamiltonian in terms of left-movers only, on an interval of length $2\ell$ with
\be H_0= iv_F\int_{-\ell}^\ell dx  \psi_-^\dagger {d\over dx}\psi_-\ee
and the  boundary condition:
\be \psi_-(\ell )=e^{2ik_F\ell}\psi_-(-\ell ).\ee
Letting
\be e^{ik_F\ell}=- ie^{i\alpha /2},\ \  (\hbox{with}\  |\alpha |<\pi  )\label{adef}\ee
it is convenient to define the field $\tilde \psi_-(x)$ by:
\be \tilde \psi_-(x)\equiv e^{-i\alpha x/(2 \ell )}\psi_-(x).\ee
$\tilde \psi$ obeys the more convenient anti-periodic boundary condition:
\be \tilde \psi_-(\ell )=-\tilde \psi_-(-\ell )\ee
at the cost of introducing a chemical potential term into $H_0$. 
Henceforth we drop the cumbersome $-$ subscript and the tilde letting:
\be \tilde \psi_-\to \psi \ee
so that 
\be H_0=v_F\int_{-\ell}^\ell dx  \psi^\dagger \left[i{d\over dx}+{\alpha\over 2\ell}\right]\psi .\ee
$H_b$ now becomes:
\be H_b\approx -\tilde t_L\gamma_L [e^{i\chi /4}\psi (0)-e^{-i\chi /4}\psi ^\dagger (0)]-
\tilde t_R\gamma_R[e^{-i\chi /4}\psi (\ell )-e^{i\chi /4}\psi ^\dagger (\ell )]\label{Hb}\ee
where
\be \tilde t_{L/R}\equiv 2\sin (k_Fa)t_{L/R}\sqrt{a}.\ee

\subsection{Single S-N Junction}
We start by considering the case of infinite $\ell$ where an incoming left-moving plane wave 
interacts with the Majorana $\gamma_L$ as it passes the origin. The fermonic operators which diagonalize the Hamiltonian are of the form:
\be \Gamma_k=\phi_{Lk}\gamma_L+\int_{-\infty}^\infty dx e^{ikx}[P_k(x)\psi(x)+H_k(x)\psi^\dagger (x)].\ee
Requiring
\be [\Gamma_k,H]=v_Fk\Gamma_k\ee
gives the Bogoliubov-DeGennes (BdG) equations:
\bea iv_F\partial_xP_k(x)&=&-2\tilde t_Le^{i\chi /4}\phi_{Lk}\delta (x)\nonumber \\
iv_F\partial_xH_k(x)&=&2\tilde t_Le^{-i\chi /4}\phi_{Lk}\delta (x)\nonumber \\
\tilde t_L[e^{i\chi /4}H_k(0)-e^{-i\chi /4}P_k(0)]&=&\phi_{Lk}v_Fk.\label{BdG}
\eea
Clearly $P_k(x)$ and $H_k(x)$ are step functions whose form is fully determined by the S-matrix:
\be \left[\begin{array}{c}
P_k(-\infty)\\
H_k(-\infty)
\end{array}\right]=\left[\begin{array}{cc}
S_{PP}(k)&S_{PH}(k)\\
S_{HP}(k)&S_{HH}(k)
\end{array}\right]\left[\begin{array}{c}
P_k(\infty)\\
H_k(\infty)
\end{array}\right]
\ee
The particle-hole symmetry of the BdG equations implies
\bea S_{HH}(k)&=&S_{PP}^*(-k)\nonumber \\
S_{HP}(k)&=&S_{PH}^*(-k).
\eea
Thus we only need to solve for $S_{PP}$ and $S_{HP}$. From Eqs. (\ref{BdG}) we find:
\bea S_{HP}(k)&=&{2\tilde t_L^2e^{-i\chi /2}\over 2\tilde t_L^2+iv_F^2k}\nonumber \\
S_{PP}(k)&=&{iv_F^2k\over 2\tilde t_L^2+iv_F^2k}.\label{Svk}\eea
Note that at zero energy:
\bea  S_{HP}(0)&=&e^{-i\chi /2}\nonumber \\
S_{PP}(0)&=&0,\label{SA}\eea
corresponding to perfect Andreev reflection (or ``Andreev transmission'' in the unfolded system). 
On the other hand at sufficiently large $|k|$, $S_{HP}(k)\approx 0$ and $S_{PP}(k)\approx 1$, 
corresponding to perfect normal reflection. The crossover length scale between these two 
behaviours is seen from Eq. (\ref{Svk}) to be 
\be \xi_M\equiv {v_F^2\over \tilde t_L^2}.\label{ximni}\ee
We define this to be the ``Majorana screening cloud length'' in the non-interacting system. Beyond 
this length scale, the featureless nature of the S-matrix indicates that the MM has been ``screened''. 
See Sec. III for a further discussion of this intuitive picture from a different viewpoint that 
also applies to the interacting case.

The fact that $\xi_M\propto \tilde t_L^{-2}$ follows  from a renormalization group scaling analysis of 
$H_b$ of Eq. (\ref{Hb}), although such sophisticated methods are not really necessary for the non-interacting model. 
$\psi (0)$ has dimension 1/2 while the MM is dimensionless.  A boundary term in the Hamiltonian must have 
dimension 1, corresponding to energy.  Therefore $\tilde t_L$ is relevant, with dimension 1/2, implying 
this scaling of $\xi_M$. Note that the situation is very different for an ordinary SN junction, 
with no MM. For a long junction, $\ell \gg \xi_0$, we may again integrate out the 
excitations of the superconductor. Now the most relevant induced interaction in the normal region 
is a proximity effect pairing term.  For spinful fermions, analyzed in [\onlinecite{ACZ}], 
this is a marginal boundary interaction $[\Delta_B\psi_\uparrow (0)\psi_\downarrow (0)+h.c.]$. 
For weak tunnelling across the SN junction $\Delta_B$ is again of order $\tilde t_L^2$, like $\xi_M$. 
However, it does not introduce a characteristic length scale, being marginal and no corresponding 
crossover of the Josephson current with junction length occurs.\cite{ACZ} Rather the current 
scales as $1/\ell$ at all lengths $\gg \xi_0$.  Nonetheless, the current depends in a non-trivial way on 
$\Delta_B$, being sinusoidal for small $\Delta_B$ and being a sawtooth for a fine-tuned large 
value of $\Delta_B$.\cite{ACZ} As we will see, the behavior of the current is much 
more interesting for a topological SN junction due to the presence of this new length scale, $\xi_M$. 
For an SN junction between an ordinary superconductor and spinless electrons, the most relevant 
induced pairing term is $\psi(0)\partial_x\psi(0)$ which is irrelevant, of dimension 2. Now $\Delta_B$
{\it does} introduce a characteristic length scale, $\xi\propto \Delta_B\propto \tilde t_L$. 
Note this length scale shrinks with decreasing tunnelling, unlike $\xi_M$ which grows. 
Starting with weak tunnelling, this scaling doesn't lead to very interesting  behavior, 
since the current is sinusoidal for all junction lengths, unlike the topological case analysed here. 

\subsection{SNS Junction}
\begin{figure}
\includegraphics*[width=0.6\linewidth]{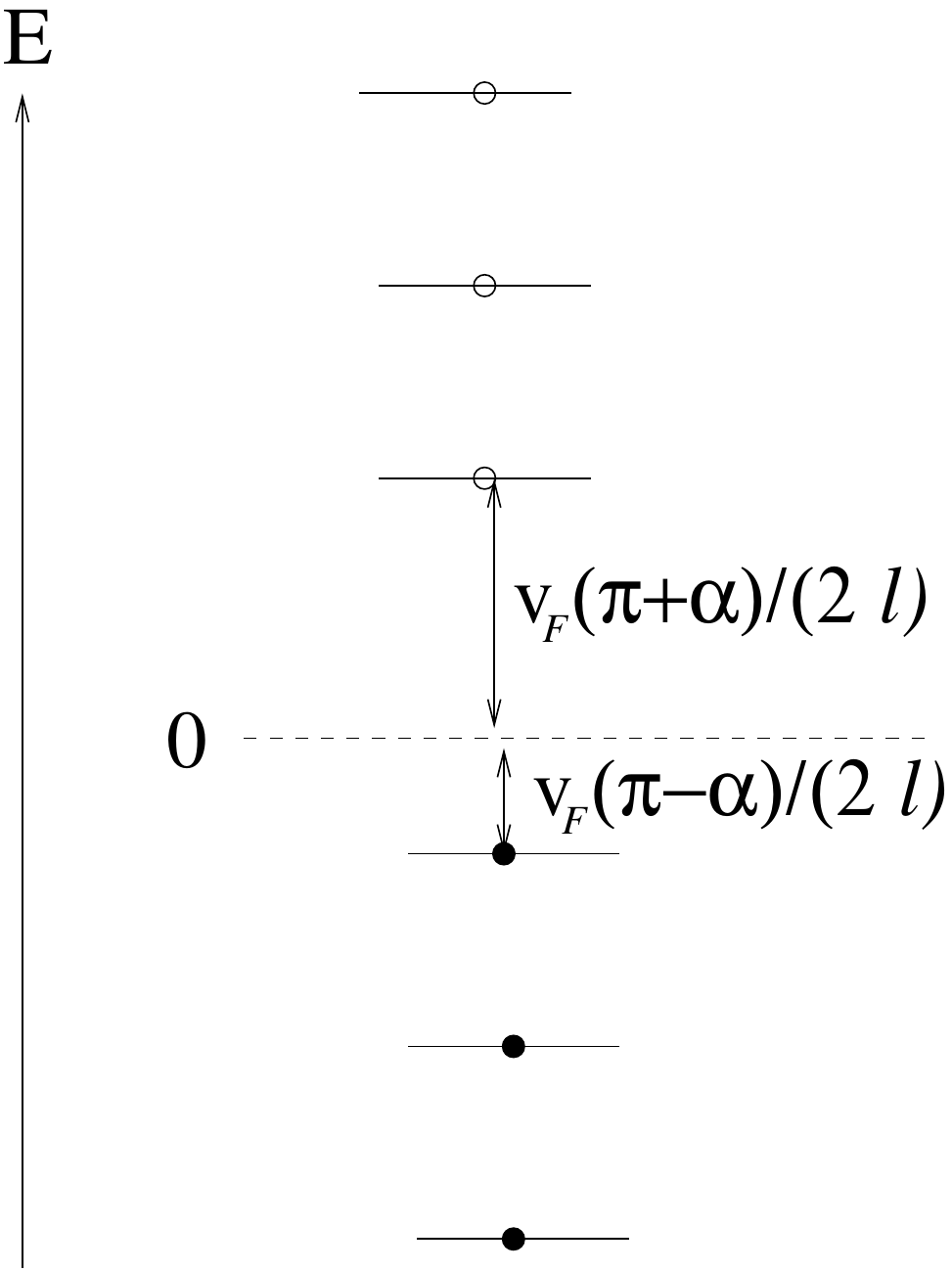}
\caption{Spectrum of Hamiltonian, with states filled in ground state marked, before coupling to MM's. Particle-hole
symmetry is broken except when $\alpha =0$ or $\pi$ and  there is a zero model when $\alpha =\pi$.}
 \label{fig:levels}
\end{figure}

In this simple non-interacting model, it is possible to write down an explicit expression for the current, for any values of 
the parameters, in terms of an elementary integral which can be readily evaluated numerically. Integrating out 
the $\psi$ field exactly gives the imaginary time effective action for the MM's:
\be S=\int_0^\infty {d\omega \over 2\pi}\left\{ {i\omega\over 2} \gamma_L (-\omega )\gamma_L(\omega )+{i\omega\over 2}
\omega \gamma_R(-\omega )\gamma_R(\omega )-
[\gamma_L(-\omega ),\gamma_R (-\omega )]{\cal M}(\omega )\left[\begin{array}{c} \gamma_L(\omega )\\ \gamma_R(\omega )\end{array}\right]
\right\}\label{S}
\ee
where
\be {\cal M}(\omega )\equiv \left[\begin{array}{cc} \tilde t_L^2[G(\omega ,x=0)-G(-\omega ,x=0)] &\tilde t_L\tilde t_R\left[e^{i\chi /2}G(\omega ,x=\ell )+
e^{-i\chi /2}G(-\omega ,x=\ell )\right] \\
-\tilde t_L\tilde t_R\left[e^{-i\chi /2}G(\omega ,x=\ell )+e^{i\chi /2}G(-\omega ,x=\ell )\right]&\tilde t_R^2[G(\omega ,x=0)-G(-\omega ,x=0)] 
\end{array}\right]\label{M}
\ee
Here $G$ is the Matsubara Green's function for the $\psi$ fermions in the case $H_b=0$:
\be G(\omega ,x )={1\over 2\ell}\sum_{n=-\infty}^\infty {e^{i\pi (2n+1)(x/2\ell )}\over i\omega -v_F[(n+1/2)\pi +\alpha /2 ]/\ell }.\label{Gwx}\ee
Note that $\alpha$ measures the amount of  breaking of particle-hole symmetry in the spectrum of $H_0$:
\be E_n={v_F\over \ell}[(n+1/2)\pi +\alpha /2].\label{fss}\ee
This spectrum is plotted in Fig. (\ref{fig:levels}).  Particle-hole symmetry is broken 
except when $|\alpha |=0$ or $\pi$  and there is  a zero mode in the spectrum when $\alpha =\pi$.
We see from Eq. (\ref{Gwx}) that

We see from Eq. (\ref{Gwx}) that
\bea G(\omega ,x=0)&=&-{1\over 2v_F}\tan \left[{i\omega \ell \over v_F}+{\alpha \over 2}\right]\nonumber \\
 G(\omega ,x=\ell )&=&{i\over 2v_F}\sec \left[{i\omega \ell \over v_F}+{\alpha \over 2}\right].\label{G0ell}
 \eea
 
\begin{figure}
\includegraphics*[width=1.\linewidth]{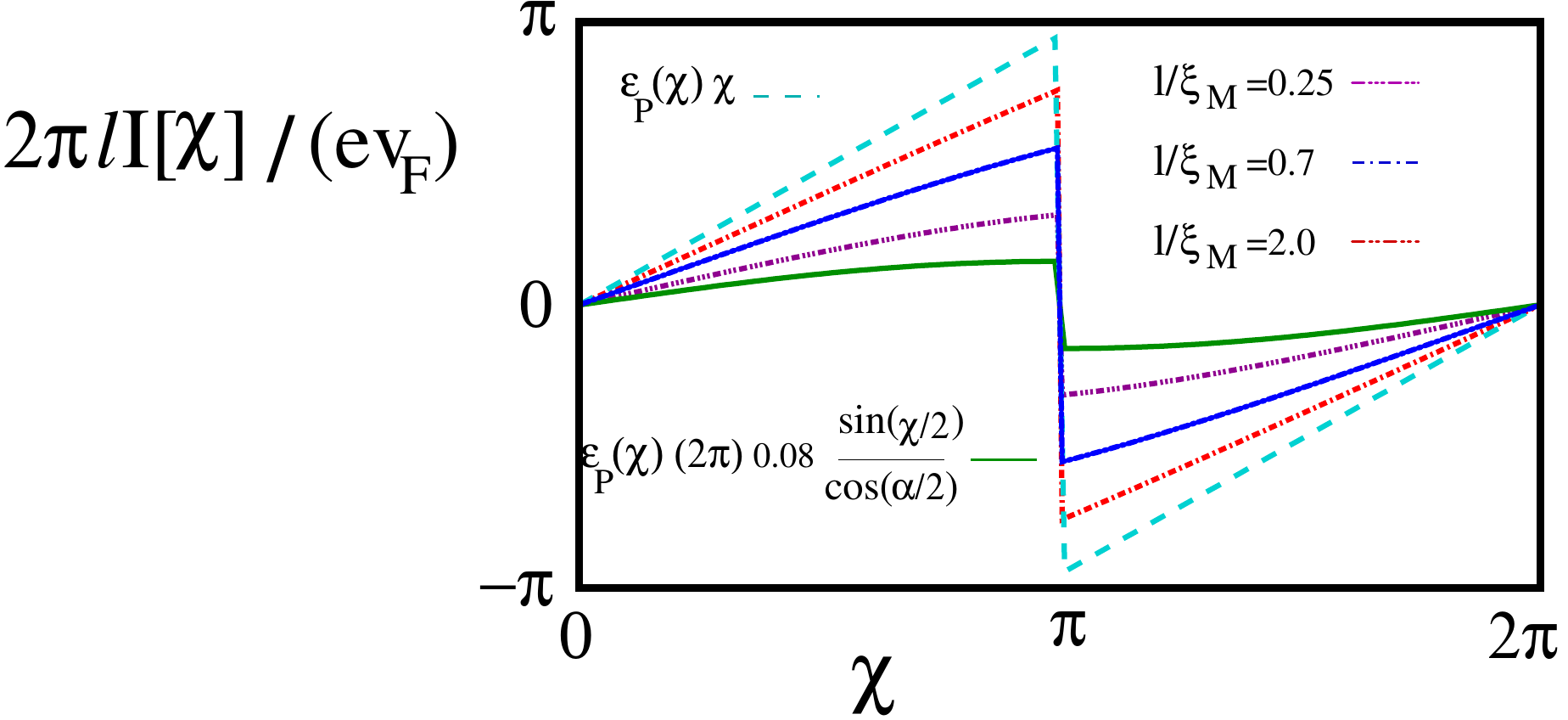}
\caption{Josephson current  versus phase difference, $\chi$ from Eq. (\ref{currex}) for particle-hole asymmetry parameter $\alpha =\pi /10$ 
and a left-right symmetric junction. $\xi_M$ is defined in Eq. (\ref{ximni}). Fermion parity conservation is {\it not} imposed. Our analytic predictions   are 
also shown, fitting the data well for $\ell \ll \xi_M$ and $\ell \gg \xi_M$. Note that steps occur for all $\ell  /\xi_M$.
}
 \label{fig:Infp}
\end{figure}

Next we integrate out the MM's to express the ground state energy as
 \be E_0=\int_0^\infty {d\omega \over 2\pi}\ln \hbox{Det} \left[ {i\omega\over 2} {\bf I}+\cal{M}(\omega )\right] .
 \ee
 Differentiating with respect to $\chi$ then gives the exact formula for the current.  In terms of a rescaled integration variable, $w=2\omega \ell /v_F$:
 \begin{eqnarray}
 I [ \chi ]&=& \frac{v_F ( 2 e) }{8 \pi \ell} \: \int_{ - \infty}^\infty \:d w \: 
 \sin \chi [ \cos \alpha   + \cosh w]
\nonumber \\
&& / \biggl\{ \left[ \frac{ v_F^2 w}{4\ell \tilde t_L\tilde t_R} ( \cos \alpha  + 
\cosh w) + \left( \frac{\tilde t_L^2+\tilde t_R^2}{2\tilde t_L\tilde t_R} \right) \sinh w \right]^2 
\nonumber \\
&-& \left( \frac{\tilde t_L^2 - \tilde t_R^2}{2\tilde t_L\tilde t_R} \right)^2 \sinh^2 w + 
  [ 1 + \cos  \alpha  \cosh w + 
\cos \chi   ( \cos  \alpha  + \cosh w) ] \biggr\}.
\label{currex}
\end{eqnarray}
Clearly $\ell I/(ev_F)$ is a scaling function of $\chi$, $\alpha$ and 
\bea z&\equiv& \ell/\sqrt{\xi_{ML}\xi_{MR}}={\ell \tilde t_L\tilde t_R\over v_F^2}\nonumber \\
r&\equiv& {\xi_{ML}\over \xi_{MR}}={\tilde t_R^2\over \tilde t_L^2}.
\eea
Note that the current depends on $k_F$ in two distinct ways. There is a weak dependence 
via the Fermi velocity, $v_F$ and then an additional, much stronger dependence via $\alpha$, the 
fractional part of $k_F\ell /\pi$. 
In Fig. (\ref{fig:Infp}) we plot the result of a numerical integration of Eq. (\ref{currex}) for 
three values of the parameters. As shown in Fig. (\ref{fig:Infp}), in the two limits $\ell /\xi_M\gg 1$ 
and $\ell /\xi_M\ll 1$ the  current is given by simple analytic expressions which can 
be obtained straightforwardly from Eq. (\ref{currex}). It is 
instructive to derive these expressions by physical arguments, given in the remainder of this section. 
These expressions are then extended to the interacting case in Sec. III.
 
\subsection{$\ell \ll \xi_M$}
In this case, for general values of $\alpha$, it is convenient to integrate out the $\psi$ field to obtain an effective action for the MM's as done above. Since the energy level spacing 
in the normal wire is $v_F/\ell$ we may ignore retardation effects in this short junction limit, $\tilde t^2/v_F\ll v_F/\ell$.
 Evaluating the Green's function at $\omega =0$, the action of Eq. (\ref{S}) reduces to a simple Hamiltonian:
\be H_{eff}=-2\tilde t_L\tilde t_R \cos (\chi /2)G(\omega =0,\ell )\gamma_L\gamma_R\label{He}\ee
where $G(\omega =0,\ell )=i/[2v_F\cos (\alpha /2)]$, from Eq. (\ref{G0ell}). 
Clearly this approach fails when $\alpha =\pi$. This is due to the presence of a zero mode in the spectrum 
of the normal region for $\tilde t_{L/R}=0$, Eq. (\ref{fss}).  A separate treatment of this special case is given in Sec. V. 
On the other hand, the exact result of Eq. (\ref{currex}) can  be used to calculate the current even in this case. 

Once we have reduced the Hamiltonian to Eq. (\ref{He}), containing only the 2 MM's, it is easy to calculate the 
Josephson current. We may combine the two MM's to make a single Dirac operator
\be \psi \equiv (\gamma_R+i\gamma_L)/2\ee
in terms of which
\be H_{eff}=2{\tilde t_L\tilde t_R\over v_F\cos \alpha /2}\cos (\chi /2)(\psi^\dagger \psi -1/2).\label{hesl}
\ee
We see that the ground state has the single fermionic energy level vacant for 
$\cos \chi /2>0$ and occupied  for $\cos \chi /2<0$:
\be E_0=-{\tilde t_L\tilde t_R\over v_F\cos \alpha /2}|\cos \chi /2 |.\ee
The corresponding Josephson current is
\be I[\chi ]=2e{\partial E_0\over \partial \chi}\approx e{\tilde t_L\tilde t_R\over v_F\cos \alpha /2}\sin (\chi /2) \varepsilon_P(\chi )
={ev_F\over \xi_M\cos (\alpha /2)}\sin (\chi /2)  \varepsilon_P(\chi ),
\label{Isl}\ee
Here $\varepsilon_P(\chi )$ is a periodic step function with period $4\pi$:
\bea \varepsilon_P(\chi )&=&1,\ \  (-\pi <\chi <\pi)\nonumber \\
&=&-1,\ \  (\pi <\chi <3\pi )\nonumber \\
&=&1,\ \  (3\pi <\chi <5\pi )\label{epP}\eea
et cetera.  Note that while neither $\sin (\chi /2)$ nor $\varepsilon_P(\chi )$ has $2\pi$ periodicity, their product does. 
Also note that we have assumed that the dominant $\chi$ dependence of the ground state energy, for $\ell \ll \xi_M$, is 
given by the lowest energy level. The other energy levels are $O(v_F/\ell )$, much larger than $\tilde t_L\tilde t_R\over v_F$
for $\ell \ll \xi_M$. They are only weakly perturbed by the coupling to the MM's and are thus insensitive 
to the phase of that coupling. 
 Eq. (\ref{Isl}) agrees well with 
the exact result of Eq. (\ref{currex}), for a small $\ell/\xi_M=.08$ plotted in Fig. (\ref{fig:Infp}).

\subsection{$\ell \gg \xi_M$}

Let's now consider the Josephson current for a long but finite length normal region with $\ell \gg \xi_M$. 
We may now write the operators which diagonalize the Hamiltonian as:
\be \Gamma_k=\phi_{Lk}\gamma_L+\phi_{Rk}\gamma_R+
\int_{-\ell}^\ell dx \left[ e^{i(k-\alpha /2\ell )x}P_k(x)\psi(x)+e^{i(k+\alpha /2\ell )x}H_k(x)\psi^\dagger (x)\right] .\ee
We can again satisfy $[H,\Gamma_k]=v_Fk\Gamma_k$ 
by solving the BdG Eqs. (\ref{BdG}) and the corresponding equations at $x=\ell$, yielding Eq. (\ref{currex}). 
Here we just discuss the low 
energy solutions with $|k|\ll 1/\xi_M$. In this limit we may approximate the transmission process at 
each interface as being purely Andreev. Thus a particle travelling to the left towards the interface at $x=0$ 
is transmitted as a hole while picking up a phase $e^{-i\chi /2}$ as we see from Eq. (\ref{SA}). 
At the other interface at $x=\ell$ it turns back into a particle picking up a further phase of $e^{-i\chi /2}$.
In addition it acquires a ``transmission phase'' of $e^{-2ik\ell}$. In this small $k$ approximation, $\alpha$, 
related to the fractional part of $2k_F\ell$, does not enter into the acquired phase due to a cancellation of particle 
and hole contributions. Imposing antiperiodic boundary conditions then determines the allowed wave-vectors by:
\be e^{-2ik\ell -i\chi }=-1.\ee
On the other hand, a hole travelling towards the interface at $x=0$ acquires a phase $e^{i\chi /2}$.  Therefore
the two types of low energy solutions correspond to wave-vectors
\be k_{n\pm}={1 \over \ell}[\pi (n+1/2)\pm \chi /2],\ \  (n\in Z)\label{k} \ee
and corresponding energies $E=v_Fk$. Remarkably, the low energy eigenvalues are independent of $\alpha$ 
for $\ell \gg 1/\xi_M$, quite unlike the opposite limit $\ell \ll \xi_M$, discussed in the previous 
sub-section, where there is very strong dependence on $\alpha$.

 Note that in this case important contributions are made to the current from many energy levels and we 
 must sum over all of them. 
 We now consider the entire set of  energy levels in a general microscopic theory 
 such as the tight-binding model introduced above.  In general the Hamiltonian can be diagonalized in the form:
 \be H=\sum_n\epsilon_n(\psi_n^\dagger \psi_n-1/2)+C\ee
 with all $\epsilon_n\geq 0$. (The constant $C$ is independent of $\chi$.) Thus the ground state energy is 
 \be E_0=-{1\over 2}\sum_n\epsilon_n+C.\ee
 The allowed wave-vectors are quite generally
 \be k_{n\pm}={\pi n+\delta_{n\pm} \over \ell}\ee
 where $k$ is now the full wave-vector, unshifted by $k_F$ as in the field theory treatment and $\delta_{n\pm}$ are phase shifts.  
  A general technique for doing such sums at large $\ell$ was developed in [\onlinecite{ACZ}]. 
To order $1/\ell$, the ground state energy can be written
 \be E_0=\ell \int_0^{k_F}{dk\over 2\pi}\epsilon (k)+{1\over 2\pi}\int_{\epsilon_0}^{\epsilon_F}d\epsilon [\delta _+(\epsilon )+\delta _-(\epsilon )]
 +{\pi v_F\over 4\ell}\left[\left(\delta_+(k_F)\over \pi\right)^2+\left(\delta_-(k_F)\over \pi\right)^2-{1\over 6}\right]+C.
 \ee
 Here $\epsilon (k)$ is the full dispersion relation of the microscopic model, $\epsilon_0$ is the bottom of the band and $\epsilon_F$
 is the Fermi energy. The reason that the $O(1/\ell)$ term arises only at $\epsilon_F$, not the bottom of the band, $\epsilon_0$, 
is related to the vanishing of $d\epsilon /dk$ at the bottom of the band, true for general dispersion relations. 

Generally, $\delta_+(\epsilon )+\delta_-(\epsilon )$ is independent of $\chi$, so only the last term, 
 of order $1/\ell$ contributes to the current.  From Eq. (\ref{k}) we see that:
 \be \delta_\pm (k_F)=(\pi \pm \chi )/2\ee
 implying:
 \be E_0(\chi )=\hbox{constant}+{v_F\chi^2\over 8\pi \ell},\ \  (\hbox{mod}\ 2\pi ).
 \ee
 and a Josephson current
 \be I={ev_F\over 2\pi\ell}\chi ,\ \  (\hbox{mod}\ 2\pi ).\label{Ill}\ee
We have determined the current for all $\chi$ by demanding $2\pi$ periodicity. We now obtain a sawtooth with jumps at 
 $\chi =(2n+1)\pi$. As we see from Eq.  (\ref{k}), these correspond to the values of $\chi$ at which an energy level passes through zero. We see from Fig. (\ref{fig:Infp}) that Eq. (\ref{Ill}) agrees well with the exact result of Eq. (\ref{currex}) 
when $\ell \gg \xi_M$.

\section{Interacting Case}
We now consider adding general interactions in the normal region.  To the tight-binding model of Eq. (\ref{Hni}) we 
could add nearest neighbor repulsive interactions between the electrons on sites $1$ to $N-1$. To the 
Hamiltonian density of Eq. (\ref{Hnic}) we could add a $\psi^\dagger_+\psi_+\psi^\dagger_-\psi_-$ term. 
(We assume that Umklapp scattering is not relevant due to an incommensurate filling factor so the 
system remains gapless.) In the low energy effective field theory approach 
it is very convenient to use ``bosonization'' techniques which map interacting fermions in one dimension 
 into non-interacting  bosons. 
The left and right movers are represented as:
\be \psi_{\mp}(x)=C\Gamma \exp \{ i\sqrt{\pi}[\phi (x)/\sqrt{K}\pm \sqrt{K}\theta (x)]\} .\ee
Here $\Gamma$ is a ``Klein factor'', another unphysical Majorana fermion introduced to obtain the 
correct anti-commutation relations with the physical MM's, $\gamma_{L/R}$ \cite{beri}.  $K$ is the Luttinger parameter, having the value $K=1$ for non-interacting fermions 
and $K<1$ for repulsive interactions. $C$ is some cut-off dependent constant which we can assume is positive. 
The low energy Hamiltonian is:
\be H_0={u\over 2}\int_0^\ell \left[\left({d\phi \over dx}\right)^2+\left({d\theta \over dx}\right)^2\right]
\ee
with the fields $\phi$ and $\theta$ obeying the canonical commutation relations:
\be [\phi (x),\theta (y)]=-{i\over 2}\hbox{sign}(x-y).\label{cr}\ee
Here $u$ is a renormalized Fermi velocity.  The boundary conditions of Eq. (\ref{bc}) become:
\bea\theta (0)&=&\sqrt{\pi /4K} \label{bc0}\\
\theta (\ell )&=&\alpha /\sqrt{4\pi K}.
\eea
$H_b$ thus becomes, in the low energy effective Hamiltonian:
\be H_b\approx -i\tilde t_L\gamma_L\Gamma Ce^{i[\sqrt{\pi/K}\phi (0)+\chi /4]}
-\tilde t_R\gamma_R\Gamma Ce^{i[\sqrt{\pi/K}\phi (\ell )-\chi /4]}+h.c.\label{Hbi}
\ee

\subsection{Single S-N Junction}
Consider first the case of an infinite normal region so that we may consider the effect of a single S-N 
junction, say the left one, with
\be H_b=-i2\tilde t_L\gamma_L\Gamma \cos [\sqrt{\pi/K}\phi (0)+\chi /4].\ee
Then this boundary operator has a renormalization group scaling dimension of 
\be d=1/(2K)\label{d}\ee
and $\tilde t_L$ is relevant for $d<1$ which includes a large range of repulsive interaction strengths, $K>1/2$. It is 
expected to renormalize to a strong couplng fixed point. 
It is natural that 
there are actually two equivalent fixed points 
at which $i<\gamma_L\Gamma>=\pm 1$. We may form a Dirac fermion from the physical MM $\gamma_L$
of the topologicial superconductor on the left side and from the Klein factor $\Gamma$ coming 
from the normal region:
\be \psi_0 = (\gamma_L+i\Gamma )/2.\ee
Then
\be i\gamma_L\Gamma = 2\psi_0^\dagger \psi_0-1\ee
and we see that the two fixed points correspond to this energy level being occupied or empty. 
This level being filled corresponds to a ``Schroedinger cat state'' in which an electron is simultaneously 
localized at the end of the superconductor and delocalized in the normal region. 
At these two fixed points $\phi (0)$ is pinned at $-\sqrt{K/\pi}\chi /4$ or $\sqrt{\pi K}-\sqrt{K/\pi}\chi /4$
 (mod $2\sqrt{\pi K}$).
The physical meaning of this boundary condition on the normal region is perfect Andreev reflection as 
can be seen from the fact that it implies:
\be \psi_+^\dagger (0)=e^{i\chi /2}\psi_-(0).\ee
Note that pinning $\phi (0)$ means that the dual field $\theta (0)$ must fluctuate wildly 
and vice versa, due to the commutation relations of Eq. (\ref{cr}).  

A cartoon picture of this 
low energy fixed point is that the MM $\gamma_L$ has paired with another MM from the normal wire, represented 
by the Klein factor, $\Gamma$, to form a Dirac mode.  The remaining low energy degrees of freedom 
of the normal wire then simply exhibit perfect Andreev reflection. This is reminiscent of the 
cartoon picture of the strong coupling Kondo fixed point discussed in Sec. I. 

We may again estimate a ``screening cloud size'', $\xi_M$ from the RG equations. Starting with a small 
tunnelling to $\gamma_L$, we see that the renormalized dimensionless tunnelling amplitude becomes O(1)
at the length scale:
\be \xi_M\approx a\left({u\over \sqrt{a}\tilde t_L}\right)^{1\over 1-d}.\label{ximi}\ee
Here $a$ is an ultraviolet cut-off scale such as the lattice constant in the tight binding model. 
Of course for the non-interacting case where $K=1$, $d=1/2$, we recover Eq. (\ref{ximni}). 

Again, we expect that the length of the SNS junction, $\ell$, will act as an infrared 
cut-off on the growth of the renormalized couplings to the two MM's with distinct simple behaviours 
in the two limits $\ell \ll \xi_M$ and $\ell \gg \xi_M$ which we now consider.

\subsection{$\ell \ll \xi_M$}
It is convenient to decompose the boson fields into left and right movers:
\bea \phi (x)&=&\phi_-(x)+\phi_+(x)\nonumber \\
\theta (x)&=&\phi_-(x)-\phi_+(x).
\eea
We may again use the boundary condition at $x=0$  of Eq. (\ref{bc0}) to ``unfold'' the system, defining
\be \phi_-(-x)\equiv \phi_+(x)+\sqrt{\pi /4K},\ \  (0<x<\ell ).\ee
The Hamiltonian can now be written in terms of left-movers, $\phi_-(x)$ only on an interval of length $2\ell$:
\be H_0=u\int_{-\ell}^\ell \left({d\phi_-\over dx}\right)^2\label{H-}\ee
with commutation relations:
\be [\phi_- (x),\phi_- (y)]=-{i\over 4}\hbox{sign}(x-y)\label{crL}\ee
and twisted boundary condition:
\be \phi_-(\ell )=\phi_-(-\ell )+\alpha /\sqrt{4\pi K},\ \  \left(\hbox{mod}\ \sqrt{\pi\over K}\right).
\ee
It is then convenient to define a new field obeying periodic  boundary conditions:
\bea \tilde \phi_-(x)&\equiv& \phi_-(x)-{\alpha x\over 2\ell \sqrt{4\pi K}}   \nonumber \\
\tilde \phi_-(\ell )&=&\tilde \phi_-(-\ell ),\ \ \left(\hbox{mod}\ \sqrt{\pi\over K}\right).
\eea
We now have:
\be H_0=u\int_{-\ell}^\ell dx \left[\left({d\tilde \phi_-\over dx}\right)^2+{\alpha \over \ell \sqrt{4\pi K}}{d\tilde \phi_-\over dx}\right].\label{Hal}\ee
The boundary interactions become:
\be H_b=-\tilde t_L\gamma_L\Gamma Ce^{i(\pi /2)(1-1/K)+i\chi /4}e^{i\sqrt{4\pi /K}\tilde \phi_-(0)}
-\tilde t_R\gamma_R\Gamma Ce^{-i\chi /4}e^{i\sqrt{4\pi /K}\tilde \phi_-(\ell )}+h.c.\ee
for a non-universal positive constant $C$. 
We may again integrate out the degrees of freedom in the central region, $\tilde \phi_-(x)$, leaving 
\be H_{eff}=-i\tilde t_L\tilde t_RC^2\gamma_L\gamma_Re^{i\chi /2-i\pi /2K}G(\omega =0,\ell )+h.c.\label{Hei}
\ee
where
\be G(\tau ,x)\equiv <e^{i\sqrt{4\pi /K}\tilde \phi_-(0,0)}e^{-i\sqrt{4\pi /K}\tilde \phi_-(\tau ,x )}>.\ee
This Green's function for a left-moving boson with periodic boundary conditions on a cylinder 
of circumference $2\ell$ may be calculated from the Green's function on the infinite plane, 
$G\propto 1/(u\tau -ix)^{1/K}$ by the conformal transformation\cite{Cardy}
\be u\tau '-ix'\to  e^{\pi (u\tau -ix)/2\ell},\ee
giving, for $\alpha =0$, and an appropriately chosen positive constant $C$, 
\be G(\tau ,x)={\pi u\over  2\ell^{1/K}\sinh [\pi (u\tau -ix)/(2\ell )]^{(1/K)}}.
\ee
Thus
\be G(\tau ,\ell )={e^{i\pi /2K}\pi u\over 2 \ell^{1/K}\cosh [\pi (u\tau )/(2\ell )]^{(1/K)}}\ee
This gives:
\be G(\omega =0,\ell )= {e^{i\pi /2K}\over \ell^{1/K-1}}\int_{-\infty}^\infty dy{1\over \cosh^{1/K}y}.\ee
The $K$-dependent phase in Eq. (\ref{Hei}) cancels and we are left with:
\be H_{eff}=-2i\tilde t_L\tilde t_RC^2\gamma_L\gamma_R \cos (\chi /2)|G(\omega =0,\ell)|.\ee
So, we see that repulsive interactions, in the case $\ell \ll \xi_M$ have the effect of suppressing 
the Josephson current by an extra factor of $1/\ell^{1/K-1}$. It now has the form
\be I\propto {e\tilde t_L\tilde t_R\over \ell^{1/K-1}}\sin \chi /2,\ \  (\hbox{mod}\ 2\pi ).\label{Iisl}\ee
This has qualitatively the same form for all values of the Luttinger parameter $K$. Note that the overall amplitude 
is proportional to
\be {\tilde t_L\tilde t_R\over \ell^{1/K-1}}\propto {1\over \ell}\left({\ell^2\over\xi_{ML}\xi_{MR}}\right)^{1-d}
\label{Iampllx}\ee
where $\xi_{ML}$ and $\xi_{MR}$ are the sizes of the screening clouds for the left and right MM's, defined in Eq. (\ref{ximi}), and 
$d=1/K$ is the scaling dimension of the tunnelling terms. 

To calculate the Green's function for non-zero $\alpha$ we must take into account the extra term in the Hamiltonian 
in Eq. (\ref{Hal}):
\be \delta H\equiv u{\alpha \over \ell \sqrt{4\pi K}}\int_{-\ell}^\ell dx {d\tilde \phi_-\over dx}\ee
which commutes with the first term in $H_0$ in Eq. (\ref{Hal}).
Writing:
\be G(\tau ,\ell )=<0|e^{i\sqrt{4\pi /K}\tilde \phi_-(0)}e^{H\tau}e^{-i\sqrt{4\pi /K}\tilde \phi_-(\ell )}e^{-H\tau}|0>
\ee
we see that the Green's function picks up an extra factor:
\be \exp \left\{-i\sqrt{4\pi /K}[\delta H,\tilde \phi_- (\ell )]\tau\right\}=\exp [-u\alpha \tau /(2K\ell )]\ee
where the commutator of Eq. (\ref{crL}) was used. Thus
\be G(\omega =0,\ell )= {e^{i\pi /2K}\over \ell^{1/K-1}}\int_{-\infty}^\infty dy{e^{-\alpha y/(\pi K)}\over \cosh^{1/K}y}
\equiv  {e^{i\pi /2K}\over \ell^{1/K-1}}f_K(\alpha ).\label{fdef}\ee
implying the current:
\be I\propto {f_K(\alpha )\tilde t_L\tilde t_R\over \ell^{1/K-1}}\sin \chi /2,\ \  (\hbox{mod}\ 2\pi ).\label{Iislga}\ee
Again, this is only valid for $|\alpha|<\pi$ 
where the integral converges. At $\alpha \to \pm \pi$ a zero mode appears in the bosonic spectrum 
which must be treated separately.  See Sec. V.

\subsection{$\ell \gg \xi_M$}
In this limit, assuming the tunnelling parameters $\tilde t_{L/R}$ in  Eq. (\ref{Hbi}) renormalize to large values, 
we expect  $\phi (0)$ and $\phi (\ell )$ to be pinned, implying that $\theta (0)$ and $\theta (\ell )$ are fluctuating. As discussed 
in sub-section 3A, this corresponds to perfect Andreev scattering. 
In this large $\ell$ limit, the finite size of the normal region doesn't prevent this renormalization from occurring independently at both boundaries. 
We may now simply forget about the MM's, which are screened and just study the free boson system with 
these Andreev boundary conditions. 
We may again transform to left-movers, $\phi_-(x)$ on an interval of length $2\ell$ with the Hamiltonian of Eq. (\ref{H-}) 
and the twisted boundary conditions, determined by Eq.  (\ref{Hbi}) :
\be \phi_-(\ell )=\phi_-(-\ell )+\sqrt{K\over \pi}{\chi \over 2}\ \  (\hbox{mod}\ \sqrt{\pi K}).\ee
The field $\phi_- (x)$ may be expanded in normal modes. 
These are harmonic oscillator modes, vanishing at $x=-\ell$ and $\ell$ as well as a soliton mode of the form:
\be \sqrt{\pi \over K}\phi_-(x)=\hbox{constant}+\left[{\chi\over 2}-\pi n\right]{x\over 2\ell}.\ee
[See Sec. V and in particular Eq. (\ref{ME}) for a more detailed discussion.]
Only the soliton mode energies have any dependence on the phase difference $\chi$ of the 
two superconductors. Thus the $\chi$ dependence of the ground state is:
\be E_0={uK\over 2\pi \ell}\left[{\chi \over 2}-\pi n\right]^2\label{Eill}\ee
with the integer $n$ chosen to minimize $E_0$:  
\be E_0={uK\chi^2\over 8\pi \ell },\ \  (\hbox{mod}\ 2\pi ).\ee
The corresponding current, $2e(dE_0/d\chi )$ is
\be I={euK\over 2\pi \ell }\chi ,\ \  (\hbox{mod}\ 2\pi ).\ee
This is again a sawtooth, as found in the non-interacting case, for $\xi_M\ll \ell$, but now the amplitude is proportional to 
 the product of velocity and Luttinger parameter $uK$. 
 
We again expect that in general $I$ can be expressed as a scaling function
\be I(\chi )={eu\over 2\pi \ell}F\left( {\ell\over \sqrt{\xi_{ML}\xi_{MR}}} ,{\xi_{ML}\over \xi_{MR}},\chi ,\alpha ,K\right). \ee
The above results are consistent with this, giving:
\bea F(z,r,\chi ,\alpha ,K) &\to& K\chi , \  \  (z\to \infty) \nonumber \\
&\propto& z^{1-1/K} f_K(\alpha )\sin \chi,\  \   (z\to 0)
\eea
where the function $f_K(\alpha )$ is defined in Eq. (\ref{fdef}).

\section{Fermion Parity}
Our treatment of the topological SNS junction so far has implicitly assumed that it is in diffusive equilibrium with a 
source of electrons at chemical potential $\mu$, which determines the parameter $\alpha$ by Eq. (\ref{adef}). 
Whether this is a valid model will depend on experimental details. If the only materials in contact with 
the normal region are superconductors and insulators then there might be a bath of Cooper pairs present 
but not of single electrons. In that case, as the phase difference $\chi$ (and other parameters) are varied 
the {\it parity} of the total number of electrons in the normal region would stay fixed, the number only 
changing in steps of two.  In that case the calculation of the Josephson current still proceeds from Eq. (\ref{Idef}) but 
the ground state energy as a function of chemical potential and other parameters must be calculated for 
a fixed fermion parity.  We again expect the Josephson current to be a universal scaling function of $\ell/\sqrt{\xi_{ML}\xi_{MR}}$ 
and other parameters, but it clearly must be a different function that was calculated above without enforcing 
fermion parity conservation. In this section we extend our calculation of the current to the fermion parity symmetric 
case, in the two simple limits $\ell \ll \xi_M$ and $\ell \gg \xi_M$ and present numerical results for intermediate 
values of this ratio.

The case of weak coupling, $\ell \ll \xi_M$ is easily dealt with since the low energy effective 
Hamiltonian of Eq. (\ref{hesl}) only contains a single Dirac energy level. Without fermion parity conservation, 
a cusp occurs in $E(\chi )$ at $|\chi |=\pi$ because the energy of this Dirac level passes through 
zero at those points, with the Dirac level switching from being occupied to empty 
in the ground state, leading to a step in the current. However, with fermion parity enforced, this 
Dirac level must stay empty (or occupied) for all $\chi$, eliminating the step in the current.
So, in this case we obtain simply:
\be I=\pm {e\tilde t_L\tilde t_R\over v_F\cos \alpha /2}\sin \chi /2,\ee
where the sign depends on whether the total number of fermions is even or odd. 
Precisely the same argument applies to the interacting case for $\ell \ll \xi_M$ so 
that the current becomes:
\be I\propto {e\tilde t_L\tilde t_R\over \ell^{1/K-1}}\sin (\chi /2),\ee
with no steps.

A similar argument can be easily constructed in the  opposite limit $\ell \gg \xi_M$, in which Eq. (\ref{Eill}) 
gives the $\chi$-dependence of the ground state energy, including interactions.  The integer $n$, which 
is the number of electrons present (measured from a convenient zero point), jumps by $\pm 1$ at $|\chi |=\pi$ 
in the ground state, leading to steps in the current. With fermion parity conservation enforced, 
$n$ may only change in units of $\pm 2$.  If the value of the fermion parity is such that $n$ must be even then
\be I={euK\over 2\pi \ell }\chi ,\ \  (\hbox{mod}\ 4\pi)\ee
 On the other 
hand, if $n$ must be odd, 
\be I={euK\over 2\pi \ell }(\chi -2\pi ),\ \  (\hbox{mod}\ 4\pi).\ee
Steps occur at $\chi=4\pi (n+1/2)$ for one value of the fermion parity quantum number 
and  at $\chi =4\pi n$ for the other. 

\begin{figure}
\includegraphics*[width=1.\linewidth]{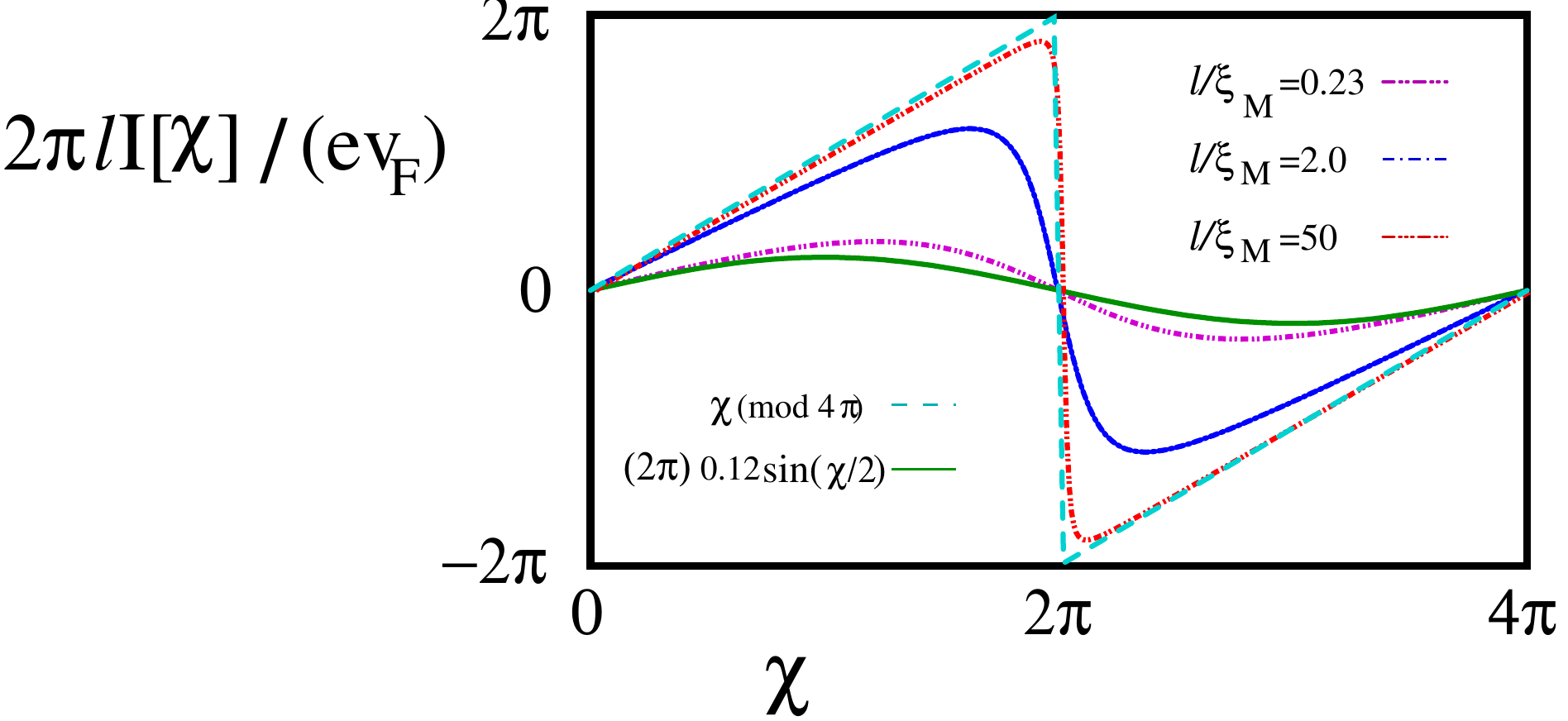}
\caption{Josephson current  versus phase difference, $\chi$ from a numerical solution of the tight binding model for particle-hole 
and left-right symmetric junction, $\alpha =0$, $\xi_{ML}=\xi_{MR}$ [defined in Eq. (\ref{ximni})] with fermion parity conservation imposed. Our analytic predictions   are 
also shown, fitting the data well for $\ell \ll \xi_M$ and $\ell \gg \xi_M$.  Note that the period is now $4\pi$ and that a step  only 
occurs at $\ell  /\xi_M\to \infty$.}
 \label{fig:Ifp}
\end{figure}

We see that the period of $I(\chi )$ is doubled from $2\pi$ to $4\pi$ when fermion parity is enforced, 
a phenomena which holds for any value of $\xi_M/\ell$. On the other hand, while 
the steps in the current are eliminated by imposing fermion parity at $\ell \ll \xi$, 
they are still present at $\ell \gg \xi$,  occurring at only 
half as many values of $\chi$ but with twice the height.  In Fig. (\ref{fig:Ifp}) we plot
the current calculated in the tight-binding model at large $N$ with fermion parity enforced, 
corresponding to $\alpha =0$ and $\xi_{ML}=\xi_{MR}$, comparing to our analytic predictions. 
(See the Appendix for a discussion of the tight-binding model results.)

\section{Zero mode case}
In the special case $|\alpha |=\pi$, the spectrum of the non-interacting normal region, Eq. (\ref{fss})  
contains a zero energy state, before 
it is coupled to the MM's.  While our analysis of $\ell \gg \xi_M$ is unaffected, 
our simple analysis of weak coupling to the MM's, $\ell \ll \xi_M$, then requires  
 modification. We may simply project the bulk fermion operators $\psi (x)$ appearing in $H_b$ onto 
the zero mode:
\be \psi (0),\psi (\ell )\to {1\over \sqrt{2\ell}}c\ee
where $c$ is the zero mode annihilation operator and the normalization of the zero-mode wave-function 
has been taken into account.  Integrating out the other, finite energy, modes, leads to no contribution to 
the low energy effective Hamiltonian to order $\tilde t_L\tilde t_R$. This can be seen from the Green's function of Eq. (\ref{Gwx}) 
at $\alpha =-\pi$ 
with the zero mode omitted:
\be G'(\omega =0,\ell )=-{1\over 2\pi v_F}\sum_{n\neq 0}{(-1)^n\over n}=0.\ee
The projected 
boundary Hamiltonian is:
\be H_b\to -{\tilde t_L\over \sqrt{2\ell}}\gamma_L[e^{i\chi /4}c-e^{-i\chi /4}c^\dagger ]
-{\tilde t_R\over \sqrt{2\ell}}\gamma_R[e^{-i\chi /4}c-e^{i\chi /4}c^\dagger ].\label{Hbap}\ee
This is a simple model of 2 MM's, $\gamma_L$, $\gamma_R$,  coupled to one Dirac mode, $c$, $c^\dagger$, 
 which can be easily diagonalized in terms of 2 Dirac mode, $\psi_\pm$, giving:
 \be H=\sum_{\pm}E_\pm (\chi )
 [\psi_\pm^\dagger \psi_\pm -1/2]\ee
 where the two positive energy eigenvalues are:
 \be E_\pm (\chi )\equiv {1\over \sqrt{2\ell}}\left[\tilde t_L^2+\tilde t_R^2\pm \sqrt{(\tilde t_L^2+\tilde t_R^2)^2-4\tilde t_L^2\tilde t_R^2\cos^2\chi /2}\right]^{1/2}.\label{Epm}\ee
 Thus the ground state energy is:
 \be E_0=-{1\over 2}\sum_{\pm}E_\pm \ee
 and the current is:
 \be I={e\tilde t_L^2\tilde t_R^2\sin \chi \over 4\sqrt{2\ell}\sqrt{(\tilde t_L^2+\tilde t_R^2)^2-4\tilde t_L^2\tilde t_R^2\cos^2\chi /2}}\sum_\pm \pm
  \left[\tilde t_L^2+\tilde t_R^2\mp \sqrt{(\tilde t_L^2+\tilde t_R^2)^2-4\tilde t_L^2\tilde t_R^2\cos^2\chi /2}\right]^{-1/2}.\label{Iznp}
 \ee
We see that this can be written:
\be I={2ev_F\over \ell}\left(\ell \over \sqrt{\xi_{ML}\xi_{MR}}\right)^{1/2}f(\xi_{ML}/\xi_{MR},\chi ),\ee
consistent with scaling. $I$  can again be seen to have steps at $|\chi |=\pi$ due to $E_-$ vanishing there. 

Naturally, the same result can be obtained from our exact result, Eq. (\ref {currex}), for the current in the non-interacting case, by 
taking the limit $\ell \ll \xi_M$. In this limit the integral is dominated by small $\omega$. Setting $\alpha =\pi$ and Taylor expanding the $\cosh$ 
and $\sinh$ functions, 
Eq. (\ref{currex}) gives:
\be I={e\over \pi \ell^2}\tilde t_L^2\tilde t_R^2\sin \chi \int_{-\infty}^\infty d\omega {1\over [\omega^2+(2E_+)^2][\omega^2+(2E_-)^2]}\ee
where $E_{\pm}$ are given in Eq. (\ref{Epm}). Doing this elementary integral gives back Eq. (\ref{Iznp}). 
 Note that $I$ is larger by a factor of $\sqrt{(\xi_{ML}\xi_{MR})^{1/2}/\ell}$, for $\ell \ll \sqrt{\xi_{ML}\xi_{MR}}$,  in 
the case $\alpha =\pi$ with a bulk zero mode than for generic $\alpha$.  The crossover behaviour 
for $\alpha$ near $\pi$ could also be calculated by considering the case of a low energy bulk state 
whose energy is not precisely zero. The result can be read off from  Eq. (\ref {currex}).

This result may be straightforwardly extended to the case with fermion parity conservation. The fermion parity of 
the ground state of this 2-state system reverses at $\chi =\pi$ where $E_-=0$. Thus the lowest energy state of 
fixed fermion parity has energy:
\be E_0^{FP}=-{1\over 2}[E_+\pm \varepsilon_P (\chi )E_-]\label{E00fp}
\ee
where $\varepsilon_P (\chi )$ is the period $4\pi$ step function defined in Eq. (\ref{epP}).    
the plus or minus sign in Eq. (\ref{E00fp}) is determined by whether the number of fermions is maintained at even or odd parity. 
Then the current becomes:
 \bea &&I={e\tilde t_L^2\tilde t_R^2\sin \chi \over 4\sqrt{2\ell}\sqrt{(\tilde t_L^2+\tilde t_R^2)^2-4\tilde t_L^2\tilde t_R^2\cos^2\chi /2}}\nonumber \\
 &&\left\{- \left[\tilde t_L^2+\tilde t_R^2+\sqrt{(\tilde t_L^2+\tilde t_R^2)^2-4\tilde t_L^2\tilde t_R^2\cos^2\chi /2}\right]^{-1/2}
 \pm \varepsilon_P (\chi ) \left[\tilde t_L^2+\tilde t_R^2- \sqrt{(\tilde t_L^2+\tilde t_R^2)^2-4\tilde t_L^2\tilde t_R^2\cos^2\chi /2}\right]^{-1/2}\right\}
 \label{Ifpzm}
 \eea
 which is continuous (no steps) and has period $4\pi$. [The $\pm$ sign in Eq. (\ref{Ifpzm}) is determined by whether 
 the fermion number is maintained at even or odd values.]

Finally, we may extend our analysis to the interacting case. At $\alpha =\pi$ there is  again a zero energy 
bulk state now represented in bosonized language. In general, the mode expansion for the left-moving bosonic field is:
\be \phi_- (x)=\phi_0+{x\over 4\ell}[Q-\sqrt{\pi /K}\alpha /\pi ]+\sum_{n=1}^\infty {1\over \sqrt{4\pi n}}
[e^{-i\pi nx/\ell}a_n+h.c.]\label{ME}
\ee
The first term represents the solitonic mode of the boson field. The operator $Q$ has  eigenvalues $Q=\sqrt{4\pi /K}n$ 
for integers $n$ and 
 is canonically conjugate to $\phi_0$:
\be [Q,\phi_0]=-i,\ee
necessary for $\phi_- (x)$ to obey the canonical commutation relations of Eq. (\ref{crL}).
The last term in Eq. (\ref{ME}) contains the harmonic oscillator modes, 
with corresponding annihilation operators $a_n$. For $\alpha =\pi$ we see that the  $Q=0$ and $Q=\sqrt{4\pi /K}$ soliton states
become degenerate ground states. For $\ell \ll \xi_M$ we project the boundary Hamiltonian of Eq. (\ref{Hbi}) onto these ground states of the 
solitonic mode. The operators $e^{\pm i\sqrt{4\pi /K}\phi_0}$ are raising and lowering operators between these 
two ground states:
\be e^{i\sqrt{4\pi /K}\phi_0}|->\propto |+>,\ \  e^{-i4\sqrt{\pi /K}\phi_0}|+>\propto |->\ee
We may represent them as raising and lowering operators for an effective s=1/2 spin:
\be e^{\pm i\sqrt{4\pi /K}\phi_0}\propto S^{\pm}.\ee
To first order in $\tilde t_{L/R}$ we can integrate out the oscillator degrees of freedom 
by simply taking the ground state expectation value of the exponential operator in $H_b$:
\be <0|e^{i\sqrt{4\pi /K}\phi (0)}|0>\propto S^+<0|\exp \left[i\sqrt{4\pi /K}\sum_{n=1}^\infty {1\over \sqrt{4\pi n}}[a_n+a_n^\dagger ]\right]|0>
= S^+\exp \left[{-1\over 2K}\sum_{n=1}^{\ell /a}{1\over n}\right].\ee
Note that the ground state matrix element is taken in the harmonic oscillator space but the solitonic factor remains as an operator. 
We have inserted an ultraviolet cut off on the number of oscillator modes, restricting the wave-vector 
to $k<\pi /a$, where $a$ is for example the lattice constant in the tight binding model, so the maximum $n$ is $O(\ell /a)$. Thus
\be <0|e^{i\sqrt{4\pi /K}\phi (0)}|0>\propto S^+\left({a\over \ell}\right)^{1/(2K)}.\ee
The exponent $1/(2K)$ appearing here is the RG scaling dimension of the operator $e^{i\sqrt{4\pi /K}\phi (0)}$. 
We obtain the same result for $e^{i\sqrt{4\pi /K}\phi (\ell)}$.
Then the boundary Hamiltonian, projected onto the two degenerate solitonic ground states, with the 
harmonic oscillator modes integrated out takes the form:
\be H_b\propto -{\tilde t_L\over \ell^{1/2K}}\gamma_L\Gamma [e^{i\chi /4}S^--e^{-i\chi /4}S^+]-
{\tilde t_R\over \ell^{1/2K}}\gamma_R\Gamma [e^{-i\chi /4}S^--e^{i\chi /4}S^+].\ee
Finally, we note that we can make a Jordan-Wigner-type transformation and represent the product 
of the Klein factor $\Gamma$ and the spin raising and lowering operators $S^{\pm}$ by a Dirac 
pair $c$, $c^\dagger$:
\bea \Gamma S^-&=&c\nonumber \\
\Gamma S^+&=&c^\dagger .\eea
We thus recover the boundary Hamiltonian of the non-interacting case, Eq. (\ref{Hbap}) apart 
from the different power of $1/\ell$. 
Thus the current is again given by Eq. (\ref{Iznp}) with this different power of $1/\ell$. The current therefore  
has the scaling form:
\be I={eu\over 2\pi \ell}\sqrt{\tilde t_L\tilde t_R}\ell^{1-1/2K}f(\tilde t_L/\tilde t_R,\chi )
={eu\over 2\pi \ell}(\ell /\sqrt{\xi_{ML}\xi_{MR}})^{1-d}g(\xi_{ML}/\xi_{MR},\chi )\ee
where Eqs. (\ref{d}) and (\ref{ximi}) were used. Comparing to the case $\ell \ll \xi_M$ with generic $\alpha$, 
Eq. (\ref{Iisl}),  Eq. (\ref{Iampllx}), 
we see that the current is enhanced by a factor of $(\sqrt{\xi_{ML}\xi_{MR}}/\ell )^{1-d}$ at $\alpha =\pi$ 
due to the presence of the bulk zero mode. 
The case with fermion parity conservation can be analysed as above for the non-interacting limit.

\section{Conclusions}
Quantum impurity models quite generally exhibit a characteristic crossover length scale, or 
screening cloud size, an idea which 
may have first emerged from Ken Wilson's work on the Kondo problem. 
We have studied this crossover in the dc Josephson current through a long normal wire between two topologicial superconductors, 
which is very generally a scaling function of the ratio of the screening cloud size associated 
with the Majorana mode at each SN junction to the length of the normal region.  
Analytic formulas were derived when this ratio is large or small, based on simple physical pictures. 
The crossover, in the non-interacting limit, was studied numerically and confirmed these 
simple pictures. Experimental measurements on such long topological SNS junctions could provide 
evidence for the existence of Majorana modes {\it and} of a Kondo-like screening cloud. 

\acknowledgments We would like to thank Y. Komijani and A. Tagliacozzo for helpful discussions. This research was supported in part by 
NSERC of Canada and CIfAR. 

\appendix 
\section{Current in tight binding model}
To diagonalize the Hamiltonian of Eq. (\ref{Hni}) we make a BdG transformation:

\be
\Gamma_E = \sum_{ j = 1}^{ N - 1 } \{ u_j c_j + v_j c_j^\dagger \} + 
w_L \gamma_L + w_R \gamma_R 
\:\:\:\: .
\label{naczm.2}
\ee
\noindent
On requiring that $ [ \Gamma_E , H ] = E \Gamma_E$ we obtain the BdG
equations

\begin{eqnarray}
 E u_j &=& - J \{ u_{j+1} + u_{j-1} \} - \mu u_j \nonumber \\
  E v_j &=&   J \{ v_{j+1} + v_{j-1} \} + \mu v_j 
  \:\:\:\: , 
  \label{naczm.3}
\end{eqnarray}
\noindent
for $2 < j < N -2$, supplemented by the boundary conditions

\begin{eqnarray}
 E u_1 &=& - J u_2 - \mu u_1 -i  t_L w_L e^{ i \frac{\chi}{4}} \nonumber \\
 E v_1 &=&  J v_2 + \mu v_1 - i t_L w_L e^{ - i \frac{\chi}{4}} \nonumber \\
 E w_L &=& 2 t_L i \{ u_1 e^{ - i \frac{\chi}{4}} + v_1  e^{ i \frac{\chi}{4}}  \}
 \:\:\:\: , 
 \label{naczm.4}
\end{eqnarray}
\noindent
at the left-hand boundary, and

\begin{eqnarray}
E u_{N - 1} &=& - J u_{N - 2} - \mu u_{N - 1} + t_R w_R  e^{ - i \frac{\chi}{4}} \nonumber \\
E v_{N - 1} &=&  J v_{N - 2} + \mu v_{N - 1} - t_R w_R  e^{   i \frac{\chi}{4}} \nonumber \\
E w_R &=&  2 t_R \{ u_{N - 1} e^{  i \frac{\chi}{4}} - v_{N - 1}  e^{ - i \frac{\chi}{4}}  \}
\:\:\:\: ,
\label{naczm.5}
\end{eqnarray}
\noindent
at the right-hand boundary. In view of the ''bulk'' equations in 
Eqs.(\ref{naczm.3}), we expect that a generic solution to the 
BdG equations is given by

\be
\left[ \begin{array}{c}
u_j \\ v_j         
       \end{array} \right]
  = \left[ \begin{array}{c}
A e^{ i k aj } + B e^{ - i ka j} \\ 
C e^{ - i k' aj } + D e^{ i k' aj} 
           \end{array} \right]
\;\;\;\; , 
\label{naczm.6}
\ee
\noindent
with

\be
E = - 2 J \cos ( ka) - \mu = 2 J \cos ( k'a) + \mu
\:\:\:\: . 
\label{naczm.7}
\ee
\noindent
In view of the explicit form of the solution in Eqs.(\ref{naczm.6}), Eqs.(\ref{naczm.4})
simplify to

\begin{eqnarray}
0  &=&  J u_0   -i  t_L w_L e^{ i \frac{\chi}{4}}\label{bcL1}\\
0 &=&  J v_0   + i t_L w_L e^{ - i \frac{\chi}{4}} \label{bcL2}\\
 E w_L &=& 2 t_L i \{ u_1 e^{ - i \frac{\chi}{4}} + v_1  e^{ i \frac{\chi}{4}}  \}
 \:\:\:\: , 
 \label{naczm.8}
\end{eqnarray}
\noindent
and, similarly, Eqs.(\ref{naczm.5})
simplify to

\begin{eqnarray}
0 &=&  J u_{N } + t_R w_R  e^{ - i \frac{\chi}{4}} \label{bcR1}\\
0  &=&  J v_{N } +  t_R w_R  e^{   i \frac{\chi}{4}} \label{bcR2} \\
E w_R &=& 2 t_R \{ u_{N - 1} e^{  i \frac{\chi}{4}} - v_{N - 1}  e^{ - i \frac{\chi}{4}}  \}
\:\:\:\: .
\label{naczm.9}
\end{eqnarray}
Substituting Eqs. (\ref{naczm.6}) into Eqs. (\ref{bcL1}), (\ref{bcL2}), (\ref{bcR1}) and (\ref{bcR2}) we can solve 
for $A$, $B$, $C$ and $D$ in terms of $w_L$ and $w_R$. Eqs. (\ref{naczm.8}) and (\ref{naczm.9}) then yield:
\noindent
\be 
{1\over J}\left(\begin{array}{cc}
2 t_L^2\left(-{\sin k(\ell -a)\over \sin k\ell}+{\sin k'(\ell -a)\over \sin k'\ell}\right)&
2i t_Lt_R\left(-{\sin k\over \sin k\ell }e^{-i\chi /2}-{\sin k'\over \sin k'\ell }e^{i\chi /2}\right)\\
2i t_Lt_R\left({\sin k\over \sin k\ell }e^{i\chi /2}+{\sin k'\over \sin k'\ell }e^{-i\chi /2}\right)&
2 t_R^2\left(-{\sin k(\ell -a)\over \sin k\ell}+{\sin k'(\ell -a)\over \sin k'\ell}\right)
\end{array}\right)\left(\begin{array}{cc}w_L\\w_R\end{array}\right)
=E\left(\begin{array}{cc}w_L\\w_R\end{array}\right).
\label{aleph.0}
\ee
The allowed values of $k$ are thus given by the solutions of:
\begin{eqnarray}
0 &=& \left\{ \left[  E J + 2 t_L^2 \left( \frac{\sin k ( \ell - a ) }{\sin k \ell} - 
 \frac{\sin k' ( \ell - a ) }{\sin k' \ell} \right) \right] 
 \left[  E J + 2 t_R^2 \left( \frac{\sin k ( \ell - a ) }{\sin k \ell} - 
 \frac{\sin k' ( \ell - a ) }{\sin k' \ell} \right) \right]  \right\}
 \nonumber \\
 &-& 4 t_L^2 t_R^2 \left( \frac{\sin ka}{\sin k \ell} e^{ - i \chi / 2} + 
  \frac{\sin k'a}{\sin k' \ell} e^{   i \chi / 2} \right) 
  \left( \frac{\sin ka}{\sin k \ell} e^{   i \chi / 2} + 
  \frac{\sin k'a}{\sin k' \ell} e^{  - i \chi / 2} \right) 
  \:\:\:\: 
  \label{aleph.1}
\end{eqnarray}
with $k$ and $k'$ related by Eq. (\ref{naczm.7}). 
A simple special case occurs at half-filling, $\mu =0$ when $k'=\pi -k$. 
In this case, Eq. (\ref{adef}) determines $\alpha =0$ for odd $N$ and $\alpha =\pi$ for even $N$. 
To get a simple solution for $\alpha =0$, we choose $N$ odd and 
also assume $t_L=t_R\equiv t$.  Eqs. (\ref{aleph.1}) 
simplifies to 
\be EJ\sin k\ell +4t^2\sin k(\ell -a)=\pm 2t^2\sin ka\cos \chi /2.\label{Etbm}\ee
We have solved Eq. (\ref{Etbm}) for for all allowed values of $k$ for odd $N$ up to 149 finding agreement with the universal field theory predictions as $N\to \infty$ 
and small $t_{L/R}$.  We may then calculate the ground state 
energy:
\be E_0(\chi )=-J\sum_n \cos[ k_n(\chi )a] \ee
where the sum is over all solutions of Eq. (\ref{Etbm}) with $|k|<\pi /2$. The Josephson current is then given by $I=2e(dE_0/d\chi )$.
To enforce fermion parity conservation, we sum over an 
even number of values of $k_n$, which means omitting the smallest negative term for a range of $\chi$. This result 
was used to produce Fig. (\ref{fig:Ifp}).


\begin{thebibliography}{99}
\bibitem{MMs} We assume the superconducting wires are long enough 
that we only need consider two MM's, one at each SN junction, which are coupled to the 
normal wire but not directly to each other. In the case of two distinct superconductors 
this will be true if the length of each superconductor is long compared to the proximity effect-induced 
coherence length, in which case the coupling to the MM's at the far ends of the superconductors 
will be exponentially small. In the case where the normal wire is contacted at both 
ends of the same topological superconductor, the two MM's which we consider have an additional coupling 
through the superconductor. However, this will be exponentially small when the 
superconductor is long compared to the coherence length.
\end{thebibliography}
\end{document}